\documentclass[12pt]{article}
\usepackage[utf8]{inputenc}
\usepackage{setspace}
\usepackage{amsmath}
\usepackage{graphics}
\usepackage{multicol}
\usepackage{subcaption}
\usepackage{graphicx}
\usepackage{url}
\usepackage{placeins}
\usepackage{amsmath}
\usepackage{amssymb}
\usepackage{pdfpages}
\usepackage{bbm}
\usepackage{graphicx}
\usepackage{xcolor}
\usepackage[labelfont=bf,tableposition=top]{caption}
\usepackage{blindtext}
\usepackage{tcolorbox}
\usepackage{amsthm}
\usepackage{lipsum}
\usepackage{multirow}
\usepackage{tabularray}
\usepackage{floatrow}
\usepackage{natbib}
\usepackage{mathptmx}

\title{Multinomial thresholded LASSO for interpretable dimension reduction of human activity sequences}
\author{Zuofu Huang, Yingling Fan, James Hodges, Julian Wolfson}
\date{}

\begin{document}

\maketitle

\begin{abstract}
    The widespread collection of data from mobile and wearable devices has created unprecedented opportunities to study human behavior in fine temporal resolution. One common structure for such data is categorical sequences—ordered, multinomial observations across many time points. These sequences present unique statistical challenges due to their high dimensionality and complex temporal dependence, including both short- and long-term correlations. Yet, there has been relatively little methodological development focusing on principled dimension reduction specifically tailored to this type of data. In this paper, we develop and evaluate approaches to identifying "key" sequence positions which distinguish sequence types. We frame this challenge as a regression problem, introduce a variety of regularization techniques that could be applied to achieve position-based dimension reduction, and evaluate them on the motivating dataset that reflects daily time use patterns collected via a smartphone application. Results show that the thresholded LASSO, a relatively underused technique, performs better than more established methods for data with complex sequential structure. 
\end{abstract}

\section{Introduction}

In recent decades, data from phones and wearable devices has gained increasing popularity, particularly for its applications in mobile health. Such data is often multi-dimensional involving location, time and biometrics (e.g. step counts and heart rate), offering a contextually meaningful picture of participants’ behavior throughout the day. An example of a dataset motivating the methods developed in this paper is from a study of implications of COVID-19 on public transportation conducted between March and June 2021 in the Minneapolis-St.Paul metropolitan area. \citep{fan2022covid} Data was collected using Daynamica, a smartphone application that captures users’ daily activities by combining sensor-based and self-reported inputs. \citep{Daynamica} The Daynamica app segments a user's day into episodes, corresponding to \textit{activities} (periods of time where a user is staying at a particular location) and \textit{trips} (periods of time in which a user is moving between locations). The start and end times of each activity are automatically generated by the app and confirmed by the user. For this study, the Daynamica app was configured to capture 13 episode types: \textit{Bike}, \textit{Bus}, \textit{Car}, \textit{Eat out}, \textit{Education}, \textit{Home}, \textit{Leisure and recreation}, \textit{Other}, \textit{Personal business}, \textit{Rail}, \textit{Shop}, \textit{Walk} and \textit{Work}. For ease of reading, in this paper we will use the term ``activity" to refer to all episodes, regardless of whether they are technically \textit{activities} or \textit{trips}. 

The data available from the app consists of a temporally ordered sequence of fixed-duration intervals, each labeled with the user's activity type during that interval. The motivating dataset contains data on 211 people (median: 13 days/person), yielding 2892 person-days of daily activity sequences. Figure 1 shows a random sample of 100 of these sequences sorted from the start by activity. One could represent such data in many different ways; we adopt the approach described in \cite{song2021visualizing} and view each day as a vector of 1440 current activity indicators at 1-minute intervals. In this application, 1-minute intervals suffice to capture all episode transitions, as episodes with duration under 1 minute cannot be detected by the app. In other applications, a different interval duration might be appropriate.

\begin{figure}[h!]
    \centering
    \includegraphics[width = 140mm]{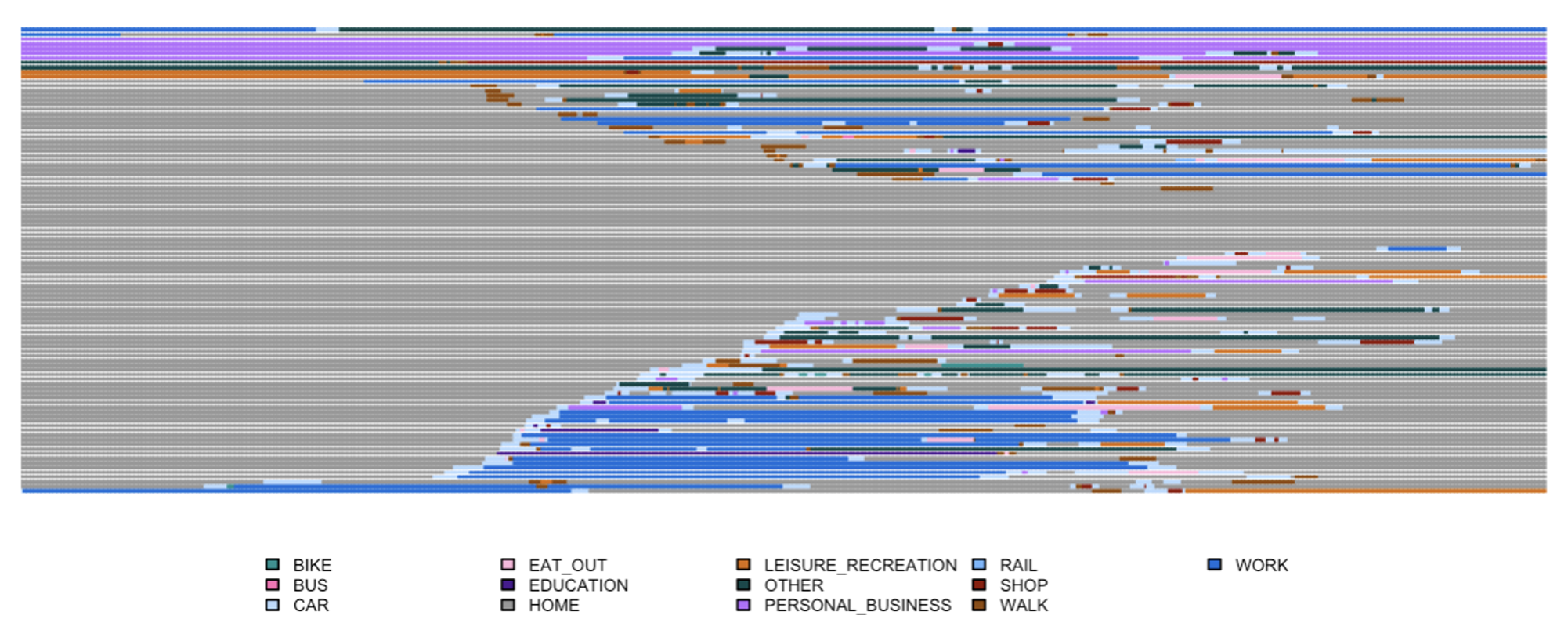}
    \caption{A sample of 100 daily sequences collected with Daynamica}
\end{figure}

\begin{figure}[h!]
    \centering
    \includegraphics[width = 120mm]{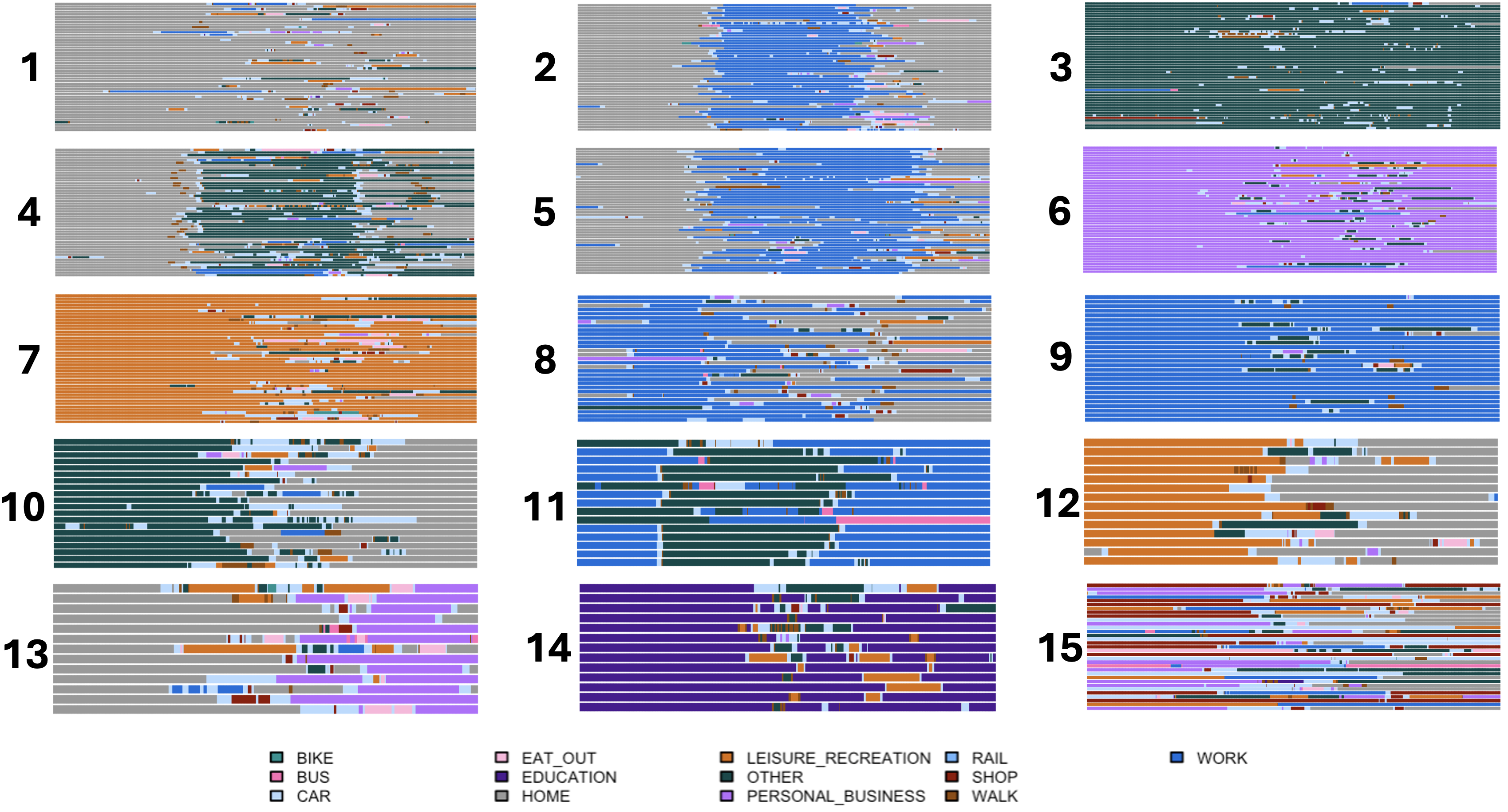}
    \caption{Sequences assigned to 15 clusters. For clusters with more than 50 sequences, a sample of 50 sequences is randomly drawn from the cluster. Cluster 15 is a collection of outlying sequences that do not belong in bigger clusters.}
    \label{data_based_assignment}
\end{figure}

The volume and complexity of such data present novel statistical challenges. Previous work has explored how to define similarity of sequences and cluster them based on shared characteristics. \citep{ben2019clustering,jiang2012clustering,song2021visualizing, barnard2025adjacency} 
For many of these methods, clusters are formed based on a distance matrix computed using sequence alignment approaches that are commonly applied for genetic sequences (see details Supplemental Materials I). Figure \ref{data_based_assignment} shows a cluster assignment for our motivating data derived using hierarchical clustering of an optimal matching distance matrix for all the day sequences. By visual inspection, it is clear that this clustering approach partitions days into groups with similar structures. A natural follow-up question is how the sequences in these clusters differ from each other. For example, in Figure \ref{data_based_assignment}, the activity types done between approximately 6 p.m. and 12 a.m. appear to be  more useful for distinguishing between clusters 2 and 5 than the activity types done between 12 a.m. and 6 a.m.  

Identifying such ``differentiating" sequence regions has benefits beyond helping to interpret clustering results. First, a set of key times/activities could yield a compressed representation of a sequence, thereby reducing data storage requirements and potentially enhancing data privacy relative to the full sequence. Further, identifying these regions may help improve study design. For example, if the goal is to study human activity patterns, it may be less useful to collect data at 3 a.m. when most people have reduced activities. Focusing intensive data collection on time periods that are most vital to determining how days differ from each other can have multiple benefits, including reduced participant burden and improved battery life of wearable devices used for data collection.

This paper explores approaches for identifying regions of a sequence that distinguish sequences from each other in terms of some characteristic of interest. In the illustration above, the characteristic of interest was a categorical cluster assignment based on the general structure of the sequences, but our methods could also be applied to discrete or continuous variables measured at the sequence level (e.g., day of the week or total daily physical activity obtained from an accelerometer). The rest of this article presents a comparison of position-based dimension reduction methods for sequential categorical data. Section 2 further motivates the problem by considering LASSO and its performance and introduces the need for thresholded LASSO. Sections 3 presents the performance of thresholded LASSO along with some other classic dimension reduction approaches. We conclude with a discussion in Section 4.

\section{Position-based dimension reduction}

\subsection{Notation and Setup}

We begin by defining some notation when episodic data of the type we have described is represented as sequences. For simplicity of notation, leave aside for now the fact that sequences (i.e., days) are often nested within individuals; we further assume that all sequences are of the same length. Suppose we have $n$ sequences available, $\{\mathbf{x}_1, \dots, \mathbf{x}_n\}$, where $\mathbf{x}_i = [x_{i1}, x_{i2}, \dots, x_{ip}]$ and $x_{ij} \sim \text{Multinomial}(\theta_{ij})$ with $\theta_{ij} \equiv \{\theta_{ijk}, k = 1, \dots, q\}$ a vector of probabilities representing the probabilities of sequence $i$ being in state $k$ at time $j$. In the data setting described above, we have $n = 2892$, $p = 1440$, and $q = 13$, though in our data some $(j,k)$ combinations are not observed for any individuals (e.g., Bike at 2 a.m.). Let $\mathbf{y} = (y_1, \dots, y_n)$ be a vector of scalar outcomes, one associated with each sequence. In this paper, we will primarily focus on the case of categorical response variables so that $y_i \sim \text{Multinomial}(\phi_i)$ with $\phi_i$ being the probability that sequence $i$ belongs in cluster $y_i \in \{1, \dots, M\}$. Section 4 briefly discusses other outcome types.

With this representation, the key challenge becomes selecting a relatively small subset of times $j$ at which variation in the states $x_{ij}$ best explains the overall structure characterized by the cluster membership.  Our problem is reminiscent of those found in the genetics and ``omics" literature, where the goal is to identify a small set of biomarkers (genes, SNPs, expressed proteins) from a long sequence which are associated with a particular phenotype. In that literature, penalized regression methods such as the LASSO are commonly used. \citep{frost2017gene, khalfaoui2018droplasso} But while human activity data share characteristics with these problems—such as their sequential nature and a large number of covariates relative to the sample size—they also differ in important ways. For instance, human activity sequences tend to display distinct state patterns and much stronger temporal dependencies. To our knowledge, little prior work has addressed these distinctive challenges in the context of human activity sequence data. For example, we hypothesized that the temporal relationship present in the data (both long term and short term to varying degrees of complexity) may present challenges for the traditional LASSO. As we discuss later, this temporal dependence is likely to violate the irrepresentability condition. \citep{zhao2006model}

Here, we focus primarily on regression-based approaches to position-based dimension reduction, assuming the generalized linear model $g[E(\mathbf{Y}|\mathbf{X})] = \mathbf{X} \beta$. Variable selection is a necessity: With a dummy variable encoding, the design matrix $\mathbf X$ has dimension $O(n \times (pq))$, and $n$ is relatively small in relation to $p$ and hence $pq$ \citep{fan2020smartphone, brown2021iterated} for example, in our motivating data $n/p \approx 2$ and $n/(pq) \approx 0.18$. 
In this regression setting, we introduce notation for coefficients that we will be referencing. Consider a multinomial regression with $15$ outcomes that estimates a list of coefficients. The length of the list is equal to 15, the number of levels in the multinomial classification in our motivating dataset. For notation purposes, we will denote the coefficient as $\beta \equiv \{\beta_{mjk}, m = 1, \dots, M; j = 1,\dots,p; k = 1, \dots, q \}$, which indicate the time points $j$ and sequence states $k$ that best explain cluster membership for each of $M$ categorical outcome levels. In many of the models below, each response level $m$ is modeled using the same structure, in which case we will use the simpler notation $\beta_{jk}$. By using each position as a predictor, the contextual meaning of each time position is naturally preserved. This preservation of context is particularly important in our setting, since the goal is to identify a specific set of positions that distinguish between clusters, and not merely to predict cluster membership using a compressed representation or derived features of the sequence. \citep{barnard2025adjacency}


\subsection{{Proposed approaches}}

We consider the following competing approaches to similarly achieve position-based dimension reduction of human activity sequences.

\subsubsection{LASSO}

Proposed by \citet{tibshirani1996regression}, the popular LASSO is defined as 
$$
\hat{\beta}^{\text{LASSO}}_m = \arg \min_{\beta} \left( -\frac{1}{2}\ell_m(\beta; \mathbf{Y}, \mathbf{X}) + \lambda \sum_{j,k} |\beta_{jk}|\right),
$$
\noindent where in our case $\ell_m$ is the logistic log-likelihood for the $m^{\text{th}}$ level of $\mathbf{Y}$. Standard packages such as \texttt{glmnet} can fit the model, although the $\lambda$ path needs to be properly tuned. The optimal $\lambda$ can be identified as the one that maximizes cross-validated prediction accuracy.

\subsubsection{Group LASSO and Sparse Group LASSO}

With the dummy variable encoding, the LASSO selects individual dummy variables corresponding to position-activity combinations (e.g., ``WORK at 9 a.m.") rather than all factors associated with the position (e.g., all 13 levels of the dummy variable at 9 a.m.). Given our goal of identifying important positions (and not position-activity combinations), it therefore seems reasonable to consider the group LASSO \citep{grouplasso} where columns of the design matrix are grouped by time to recognize their nested structure, as other activities in the position may offer ancillary information that boosts predictive power. For this setup, the group LASSO coefficient estimate for outcome level $m$ is defined by:
$$
\hat{\beta}^{GL}_m = \arg \underset{\beta}{\min} \left(-\frac{1}{2}\ell_m(\beta; \mathbf{Y}, \mathbf{X}) + \lambda \sum_{j=1}^p \sqrt{s_j}||\beta_{j}||_{2}\right),
$$

\noindent where $j$ denotes the group membership and $s_j$ is the length of $\beta_{j} \equiv \{\beta_{j k}, k=1, \dots, q\}$. The group LASSO will identify important positions, but not particular activities that are relevant at those times. In some contexts, this will be helpful, for example, to help determine specific times where data collection should be intensified. We also consider the sparse group LASSO \citep{sparsegrouplasso}  which accounts for potential sparse effects on a between-group and within-group level:
$$
\hat{\beta}^{SGL}_m = \arg \underset{\beta}{\min} \left(-\frac{1}{2}\ell_m(\beta; \mathbf{Y}, \mathbf{X}) + \lambda \left[ (1-\alpha)\sum_{j=1}^p \sqrt{s_j}||\beta_{j}||_{2} + \alpha \sum_{j,k} |\beta_{jk}| \right] \right)
$$

\noindent The R package \texttt{msgl} was used to implement the group LASSO and sparse group LASSO \citep{msgl_csda, msgl_manual}.

\subsubsection{Thresholded LASSO}

As noted previously, the predictors in our setting have strong and complex temporal dependence. It is well-known \citep{meinshausen2006high} that the Lasso has a tendency to ``over-select" variables, particularly when columns of the design matrix are strongly dependent. Therefore, we also explored regularized regression approaches that can be used to induce greater sparsity than the standard and group lasso approaches.

\citet{zhou2010thresholded} initially proposed the thresholded LASSO as a two-step procedure. In the first step, an initial estimate $\hat \beta$ is obtained and each element of $\hat \beta$ is subjected to a hard threshold $\delta$, yielding $\hat \beta^{TL}_{0} = \hat \beta \times \delta_\tau(\hat \beta)$, where for each element of $\hat \beta_j$ of $\hat \beta$,  $\delta_\tau(\hat \beta_j) = {\mathbbm{1}} [|\hat \beta_j| \geq \tau]$. The set of selected variables that survives the thresholding, $I_{TL} = \{j: \hat \beta^{TL}_{0,j} \neq 0\}$ is then used in another (regularized or unregularized) regression model to obtain final model estimates $\hat \beta^{TL}$ . For the linear model, under a variety of structural assumptions and regularity conditions, \citet{zhou2010thresholded} showed that the thresholded LASSO can identify the set of truly non-zero coefficients with probability approaching 1 at a rate substantially faster than the standard LASSO. The asymptotic analysis of the thresholded LASSO estimator suggests that, for finite sample sizes, incorporating hard-thresholding into variable selection may yield better models. \citep{zhou2010thresholded, van2011adaptive} The level of sparsity, however, is often unknown before penalized regression, which means that the thresholding scalar $\tau$ is hard to determine a priori.

To the best of our knowledge, thresholded LASSO has not been widely used in analyzing real data. We adopt it here as a viable alternative, because standard LASSO may underperform given the structure of our data. For the practical purpose of dimension reduction for human activity sequences, we propose a similar process with the goal of increased model flexibility. In our setting, we apply thresholded penalized multinomial logistic regression models instead of the linear models described in \citet{zhou2010thresholded}. The process for the thresholded LASSO is as follows:

\vspace{7mm}

{\setstretch{1.4}

\framebox{\parbox{\dimexpr\linewidth-2\fboxsep-2\fboxrule}{\itshape
{\bf{Algorithm 1: Perform thresholded LASSO}}
\begin{enumerate}
    \item Perform multinomial logistic regression with LASSO penalty using an appropriate tuning grid for $\lambda$, with the prediction accuracy as the metric for 10-fold cross validation, obtaining the coefficients $\hat \beta_{mjk}$.
    \item For a chosen threshold $t_0$ (further discussion on how to choose this threshold is provided below):
    \begin{enumerate}
        \item Obtain a set of positions $J_m = \{\beta_{mjk}: |\hat \beta_{mjk}| \geq t_0\}$ from the remaining coefficients for each level $m$.
        \item Obtain a set of positions $J =\bigcup_{m=1}^{15} J_m$. This union set represents the remaining positions actively used in the model.
        \item Perform multinomial LASSO again only using the columns of $\mathbf{X}$ corresponding to positions in $J$. Retain remaining positions with a non-zero coefficient as the final set of predictors. 
    \end{enumerate}
\end{enumerate}
}}
\label{algorithm}
}

We make two comments about the above algorithm. First, the choice of threshold in Step 2 is vital to the algorithm's performance, as there is often a trade-off between the number of remaining positions and prediction accuracy depending on the chosen threshold. It is advisable to select the threshold from a list of potential thresholds using the distribution of coefficients overall and at each level. The desired threshold should be selected given the data context and analysis goal. Second, few additional variables were dropped by LASSO at Step 2c in our application and could have been omitted without a substantial change in performance. However, it is good practice to confirm that the prediction accuracy is comparable before and after thresholding. Another reason for including it to be consistent with the original algorithm proposed by \citet{zhou2010thresholded}.

\subsubsection{Repeated LASSO}

The key to thresholded LASSO is the step of removing less ``influential" variables ranked by the magnitude of the regression coefficient. Since the hard thresholding step essentially performs additional variable selection, we could also consider an iterative process that replaces the need for hard thresholding by continuously performing variable selection. After the initial LASSO run, we repeat the process with remaining positions as predictors for the next LASSO run (detailed in Algorithm 1 Step 2(a)-2(b)), until the number of remaining predictors stabilizes (defined by no additional selection for 5 consecutive LASSO runs).

\subsubsection{Random forests}

Classification trees are another commonly used supervised learning method for prediction. Depending on the complicated between- and within-group patterns, classification trees may be better at capturing non-linear relationships in our data. Random forests aggregate results from individual trees, and individual variable importance may be determined using criteria such as the average decrease in Gini impurity \citep{breiman2001random}.

\section{Results}

In this section, we show how the methods discussed above can be used to identify positions in the sequential categorical data from Figure \ref{data_based_assignment} which best predict cluster membership.  

\subsection{LASSO}

We fit a penalized multinomial regression with LASSO penalty with the dummy variable encoded design matrix, selecting the $\lambda$ that maximizes cross-validated accuracy using the \texttt{caret} package. \citep{caret} A list of coefficients was obtained, with the length of the list equal to the number of levels in the multinomial classification. We then aggregated the results by taking the union of positions selected from each binary model, which was described in Algorithm 1 Step 2(a)-2(b).

\begin{figure}[h!]
    \begin{subfigure}{.5\textwidth}
        \centering
        \includegraphics[width = 65mm]{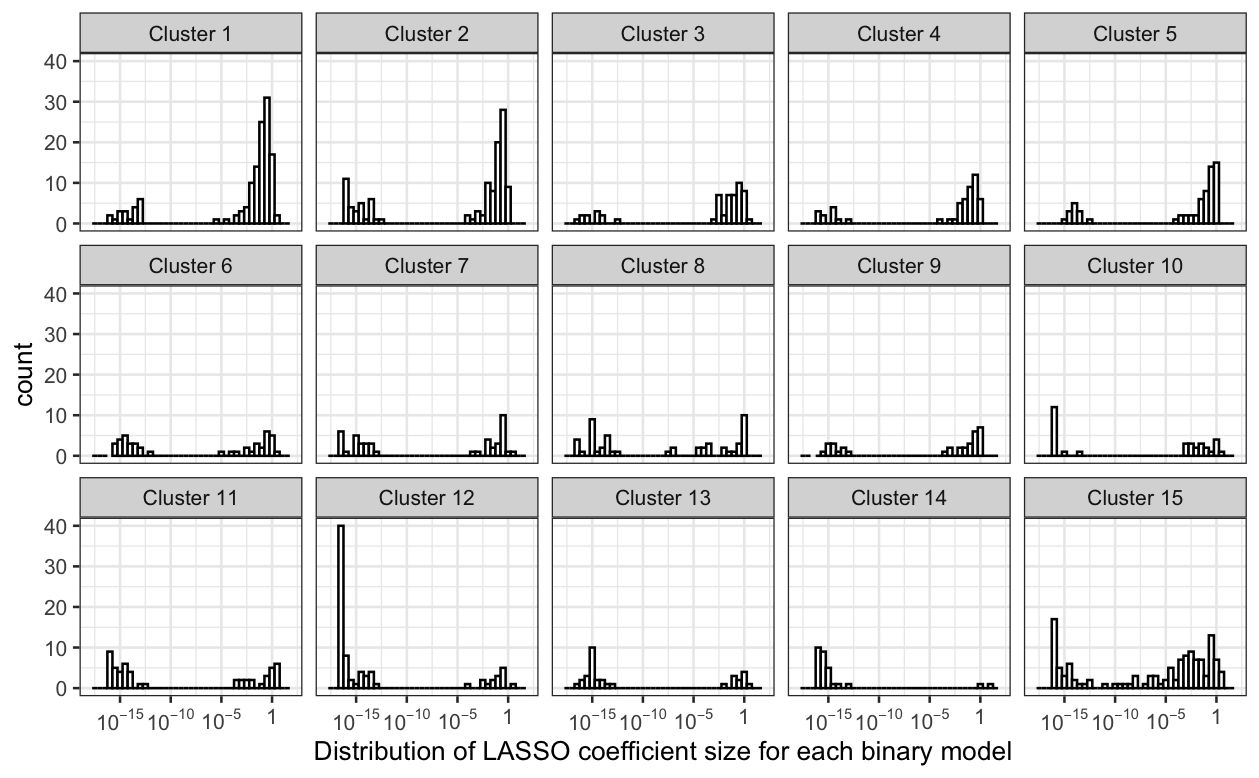}
        \caption{}
    \end{subfigure}%
    \begin{subfigure}{.5\textwidth}
        \centering
        \includegraphics[width = 65mm]{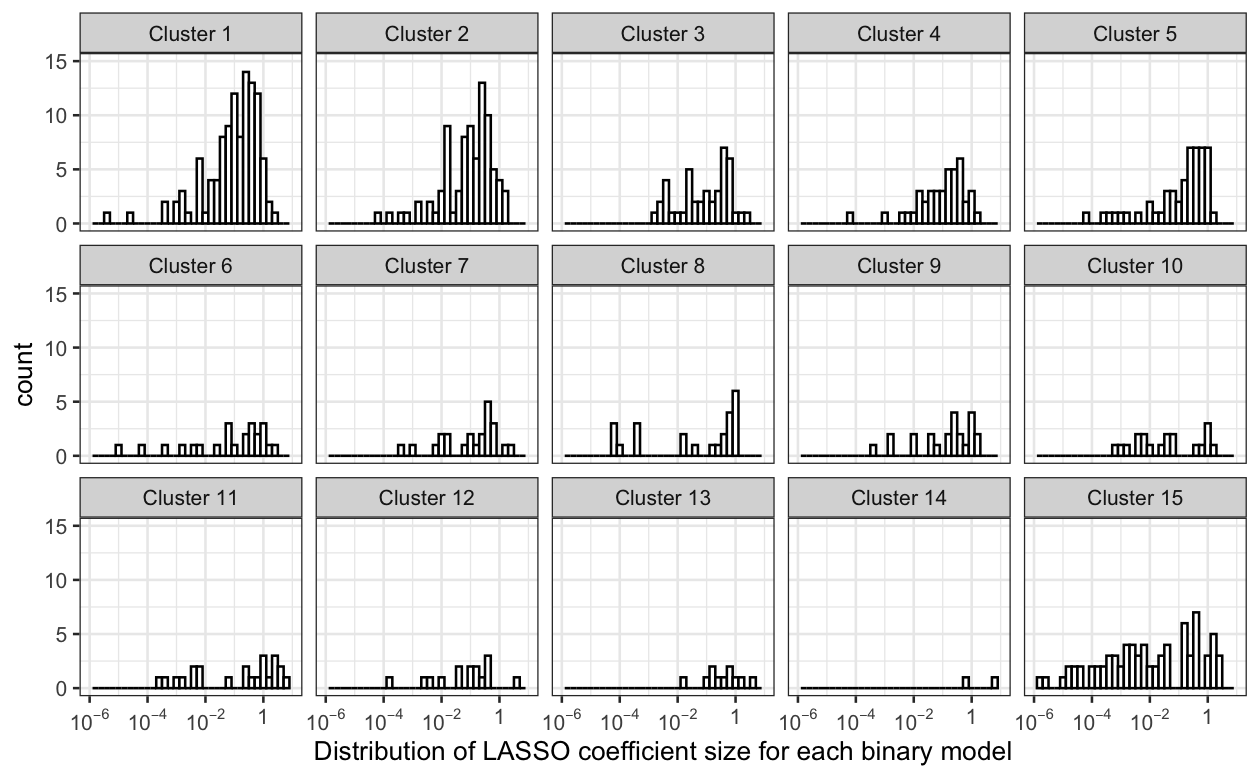}
        \caption{}
    \end{subfigure}
    \caption{Distribution of regression coefficient size for each binary model with range (a) $10^{-18}$ to $10^{2}$ and (b) $10^{-6}$ to $10$ (zoom-in)}
    \label{lasso_results}
\end{figure}

The lowest cross-validated misclassification rate of 1.4\% was achieved when 743 positions were selected, i.e., more than half of the 1,440 original positions were identified as "important", so no substantial dimension reduction was achieved. Figure \ref{lasso_results} presents the distribution of regression coefficients on a log 10 scale for each binary model. The estimated values of selected coefficients clustered around $10^{-15}$ and $10^{-2}$. Variables with coefficients clustered around $10^{-15}$ inherently have little predictive value, which suggests that potentially a more stringent threshold could be helpful to identify important predictors.

While a larger $\lambda$ shrinks more coefficients toward 0, dimensionality is not substantially reduced here. Table \ref{lasso_path_table} shows that the LASSO is on the wrong path in selecting the correct coefficients. An alternative method is warranted.

\begin{table}[h!]
\begin{tabular}{ccc}
$\lambda$ & Positions with $|\hat \beta| \geq \lambda$& Misclassification rate \\ \hline
0.0025    &  743  & 1.4\%  \\ \hline
0.00375    &  743  & 1.4\%  \\ \hline
0.005  &  601    & 1.9\%   \\ \hline
0.00625  &  548    & 2.6\%   \\ \hline
0.0075  &  535    & 3.4\%   \\ \hline
0.00875  &  449    & 4.1\%   \\ \hline
0.01  &  433    & 4.5\%   \\ \hline
0.0125  &  363    & 5.5\%   \\ \hline

\end{tabular}
\caption{Number of remaining positions predictive of the cluster assignment and misclassification rate (estimated with 10-fold CV) under various $\lambda$ values from LASSO}
\label{lasso_path_table}
\end{table}

\subsection{Thresholded LASSO}

\begin{figure}[htp]

\centering
\includegraphics[width=.33\textwidth]{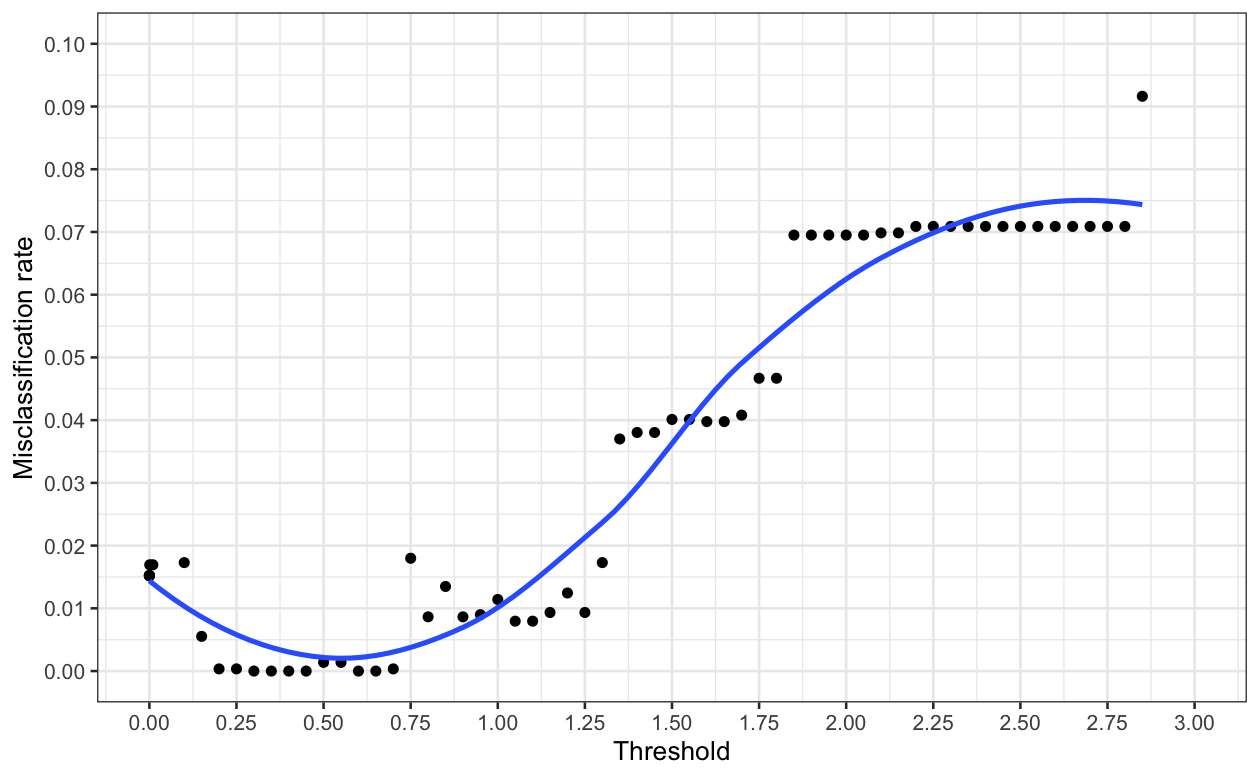}\hfill
\includegraphics[width=.33\textwidth]{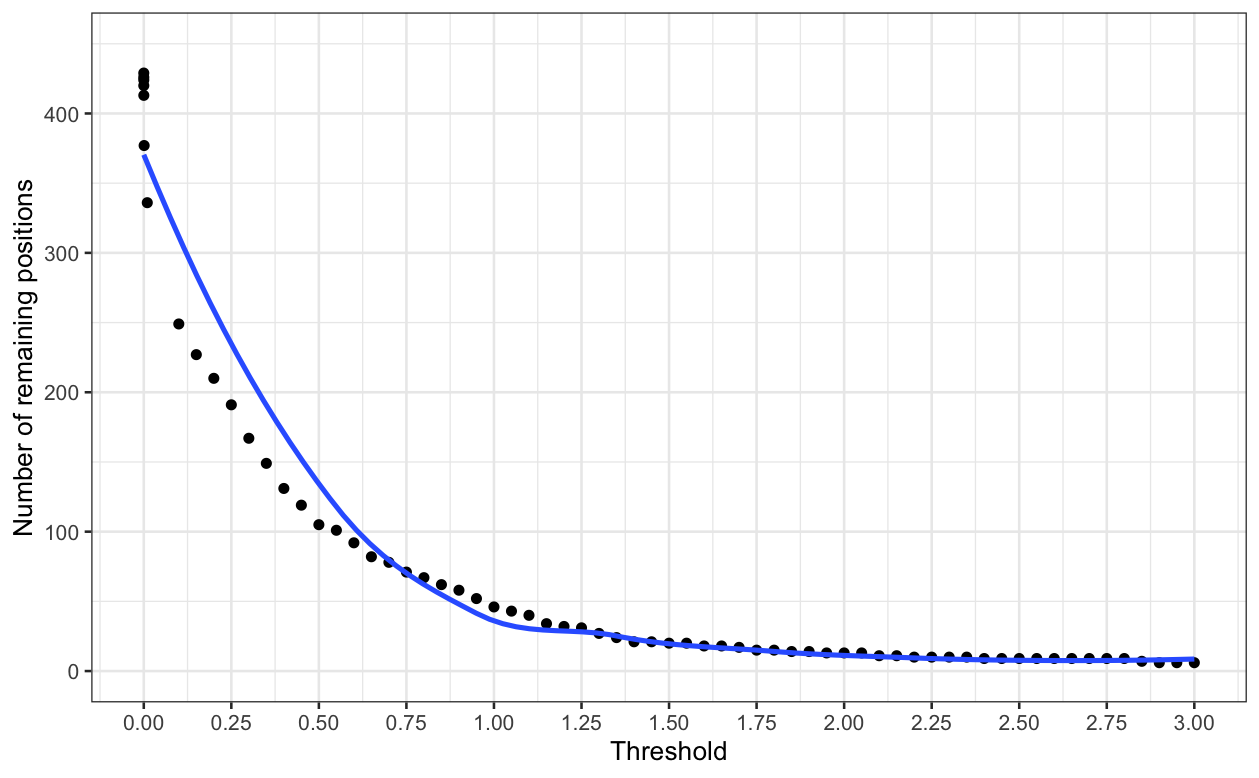}\hfill
\includegraphics[width=.33\textwidth]{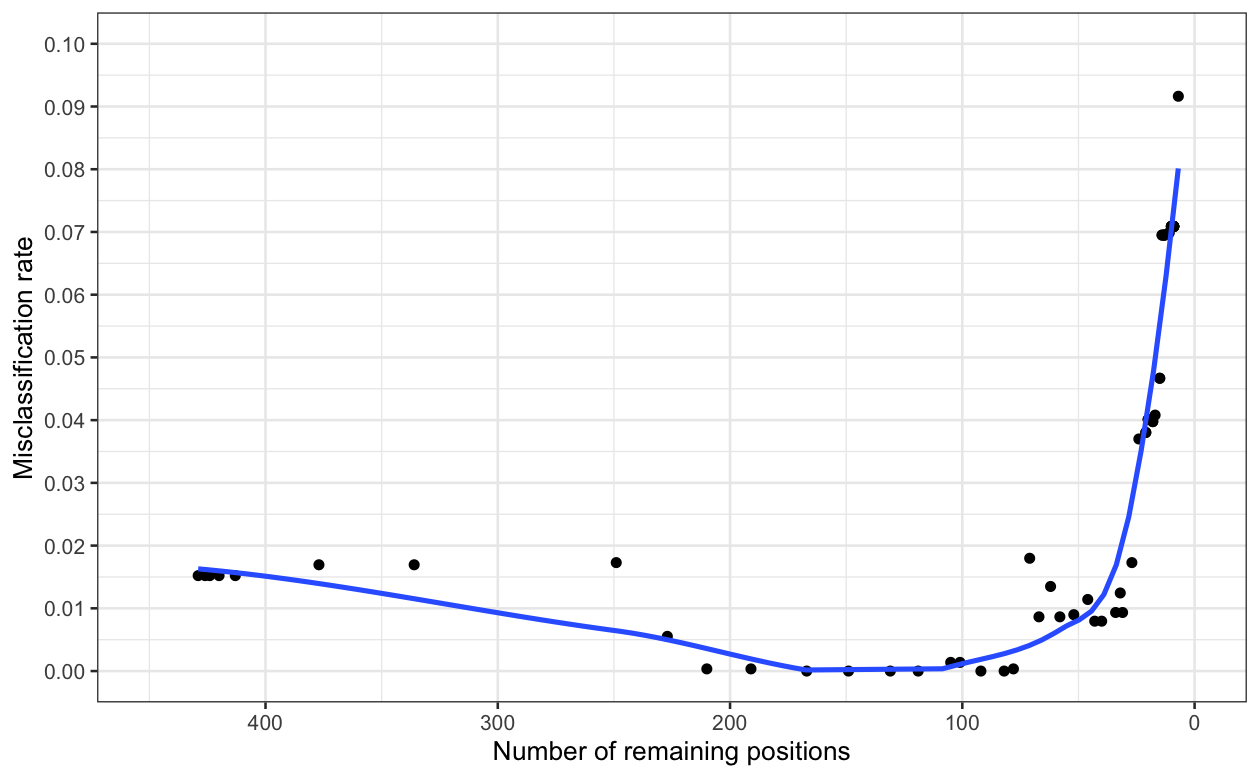}
\caption{Relationship between the hard threshold, number of remaining positions and misclassification rate for thresholded LASSO}
\label{tlasso}

\end{figure}

Figure \ref{tlasso} summarizes the number of remaining positions and the misclassification rate after applying various hard thresholds. LASSO with hard thresholding could maintain a perfect classification result for all sequences with approximately 82 out of 1440 positions and maintain a low misclassification rate, around 1\%, with approximately 30 out of 1440 positions. Too few or too many predictors result in an increased misclassification rate, but the number of predictors to keep in the model is otherwise subject to the simplicity-accuracy tradeoff.

\subsection{Group LASSO and sparse group LASSO}

Group LASSO did not perform well: Retaining 50 positions achieved a misclassification rate of 13\%. Sparse group LASSO performed similarly. Figure \ref{combined_result} presents the results for sparse group LASSO with $\alpha = 0.5$. Other $\alpha \in (0,1)$ values were also explored, with results omitted due to space constraint.

\subsection{Repeated LASSO}

A subset of 743 positions remained after the initial LASSO that uses all positions as predictors for the cluster assignment. We then repeated the process until stabilization. The number of significant predictors initially dropped but stabilized at 296 predictors after 72 repeated LASSO runs, giving a misclassification rate of 1.5\%. Compared to the thresholded LASSO, repeated LASSO made more mistakes with more predictors. This shows that a significant percentage of the variables initially selected are insensitive to additional variable selection, which justifies manual adjustment.


\subsection{Random forest}

\begin{figure}[h!]
    \centering
    \includegraphics[width = 80mm]{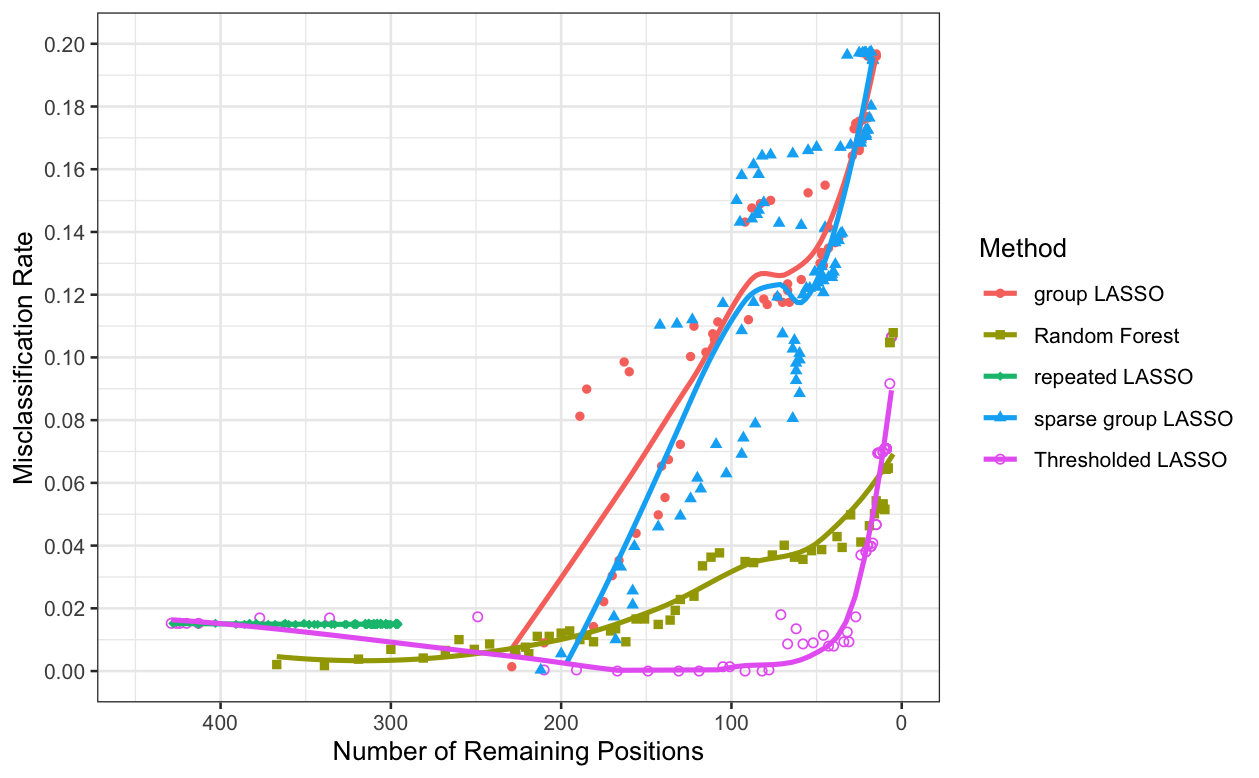}
    \caption{Number of remaining positions and corresponding misclassification rate for group LASSO, sparse group LASSO and the random forest compared to thresholded LASSO}
    \label{combined_result}
\end{figure}

For comparison purposes, we fit a random forest containing 100 recursive partitioning trees with 10-fold cross validation and maximum depth from 1 to 15. Results were aggregated from individual trees, and the average decrease in Gini impurity was used to determine and rank the importance of each activity at each position. The position was selected when the average decrease in Gini impurity for any activity associated with the position reached a certain threshold. Figure \ref{combined_result} shows that thresholded LASSO was able to achieve a higher accuracy with fewer variables compared to the random forest.

\section{Discussion}

In this paper, we describe several techniques for interpretable dimension reduction for sequential categorical data, with a motivating example of understanding how activities taking place at different times of the day explain overall similarities and differences in daily human activity sequences. We showed that, while several approaches have similar best classification accuracy, the standard LASSO and random forest identify a much larger number of positions to achieve this accuracy than the lesser-known thresholded LASSO technique. Hard thresholding in LASSO has received limited attention in applied settings. In this paper, we applied thresholded LASSO to a large-scale human activity dataset, demonstrating its utility for dimension reduction in a complex application. 

Next, we briefly reflect on why the thresholded LASSO performs superiorly to competing methods. The regular LASSO is known to over-select variables. \citep{meinshausen2006high} This is especially the case when the irrepresentability condition $\| X_2^T X_1 (X_1^T X_1)^{-1} \text{sign}(\beta_1) \|_{\infty} < 1$ is violated (where $X_1$ and $X_2$ contain columns with relevant and irrelevant variables respectively), which in essence means that relevant variables cannot be too correlated with irrelevant ones. \citep{zhao2006model} In human activity sequences, correlations among covariates (both long term and short term, temporally) are very likely. While variable selection tasks are still meaningful, the LASSO will be expected to perform poorly from a theoretical perspective. The large number of categories in the multinomial outcome may further challenge the performance of LASSO and similar methods, introducing more noise or causing LASSO to overselect variables to capture subtle differences between categories. However, this might not be the primary reason for the regular LASSO's poor performance, since overselection issues were also present in the binary models.

We note a few limitations of our work. First, thresholded LASSO may be considered a flexible process with an explicit simplicity-accuracy tradeoff. While there is no stipulated truth of what a correct threshold should be, an appropriate threshold often strikes the right balance between the number of predictors used and accuracy. However, selecting the hard threshold may be ad-hoc and depend on the data context and analysis goal. That is why we emphasize the ability to better understand the relationship between the threshold, the number of positions remaining and accuracy through a grid of possible thresholds. Second, while we present thresholded LASSO as an alternative suitable technique, many variants of the LASSO penalty exist, such as the adaptive LASSO and the smoothly clipped absolute deviation (SCAD) penalty, which we did not discuss in this paper. \citep{zou2006adaptive, fan2001variable} Some of these methods may be capable of effective variable selection in our scenario. Third, while the thresholded LASSO offers superior performance by selecting certain dummy variable levels of each position, the prediction power may be even higher when all factors levels of a position are grouped together for estimation. This also highlights an opportunity for methodological advancement at the intersection of group LASSO/sparse group LASSO and hard thresholding.

Finally, we discuss some directions for future research. First, our analysis focused on a classification setting with a multinomial outcome. Given the flexible overarching generalized linear models framework, the thresholded LASSO could be extended to continuous or other types of outcomes. The extent of performance gains in these alternative contexts warrants further investigation. Second, we hope to explore the benefit of improving study design following dimension reduction. Depending on the important positions being uncovered, less/more data (e.g. surveys of the participant's mood) may be collected at different times of the day to allow a more flexible, comprehensive understanding of the activity patterns. Lastly, we aim to extend the application of thresholded LASSO to other forms of sequential categorical data or, more broadly, to multinomially correlated data. This may include broad applications ranging from longitudinal health and transportation studies to statistical genetics problems such as gene finding. 



\section*{Software}

Software (R code) and the motivating dataset used are available on GitHub at https://github.com/zuofuhuang/reduceSeq.

\newpage

\bibliographystyle{apalike}
\bibliography{bibtex}

\begin{thebibliography}{}

\bibitem[Barnard et~al., 2025]{barnard2025adjacency}
Barnard, M., Fan, Y., and Wolfson, J. (2025).
\newblock Adjacency matrix decomposition clustering for human activity data.
\newblock {\em Journal of the American Statistical Association}, (just-accepted):1--18.

\bibitem[Ben-Gal et~al., 2019]{ben2019clustering}
Ben-Gal, I., Weinstock, S., Singer, G., and Bambos, N. (2019).
\newblock Clustering users by their mobility behavioral patterns.
\newblock {\em ACM Transactions on Knowledge Discovery from Data (TKDD)}, 13(4):1--28.

\bibitem[Breiman, 2001]{breiman2001random}
Breiman, L. (2001).
\newblock Random forests.
\newblock {\em Machine learning}, 45:5--32.

\bibitem[Brown et~al., 2021]{brown2021iterated}
Brown, R., Fan, Y., Das, K., and Wolfson, J. (2021).
\newblock Iterated multisource exchangeability models for individualized inference with an application to mobile sensor data.
\newblock {\em Biometrics}, 77(2):401--412.

\bibitem[Fan and Li, 2001]{fan2001variable}
Fan, J. and Li, R. (2001).
\newblock Variable selection via nonconcave penalized likelihood and its oracle properties.
\newblock {\em Journal of the American statistical Association}, 96(456):1348--1360.

\bibitem[Fan et~al., 2022]{fan2022covid}
Fan, Y., Becker, A., Ryan, G., and Wolfson, J. (2022).
\newblock Covid-19 implications on public transportation: Understanding post-pandemic transportation needs, behaviors, and experiences.
\newblock Research report, Center for Transportation Studies, University of Minnesota, Minneapolis, MN.

\bibitem[Fan et~al., 2020]{fan2020smartphone}
Fan, Y., Becker, A., Ryan, G., Wolfson, J., Guthrie, A., and Liao, C.-F. (2020).
\newblock Smartphone-based interventions for sustainable travel behavior: The university of minnesota parking contract holder study.

\bibitem[Fan et~al., 2015]{Daynamica}
Fan, Y., Wolfson, J., Adomavicius, G., Vardhan~Das, K., Khandelwal, Y., and Kang, J. (2015).
\newblock Smartrac: A smartphone solution for context-aware travel and activity capturing.

\bibitem[Frost and Amos, 2017]{frost2017gene}
Frost, H.~R. and Amos, C.~I. (2017).
\newblock Gene set selection via lasso penalized regression (slpr).
\newblock {\em Nucleic acids research}, 45(12):e114--e114.

\bibitem[Jiang et~al., 2012]{jiang2012clustering}
Jiang, S., Ferreira, J., and Gonz{\'a}lez, M.~C. (2012).
\newblock Clustering daily patterns of human activities in the city.
\newblock {\em Data Mining and Knowledge Discovery}, 25:478--510.

\bibitem[Khalfaoui and Vert, 2018]{khalfaoui2018droplasso}
Khalfaoui, B. and Vert, J.-P. (2018).
\newblock Droplasso: A robust variant of lasso for single cell rna-seq data.
\newblock {\em arXiv preprint arXiv:1802.09381}.

\bibitem[{Kuhn} and {Max}, 2008]{caret}
{Kuhn} and {Max} (2008).
\newblock Building predictive models in r using the caret package.
\newblock {\em Journal of Statistical Software}, 28(5):1–26.

\bibitem[Meinshausen and B{\"u}hlmann, 2006]{meinshausen2006high}
Meinshausen, N. and B{\"u}hlmann, P. (2006).
\newblock High-dimensional graphs and variable selection with the lasso.

\bibitem[Simon et~al., 2013]{sparsegrouplasso}
Simon, N., Friedman, J., Hastie, T., and Tibshirani, R. (2013).
\newblock A sparse-group lasso.
\newblock {\em Journal of computational and graphical statistics}, 22(2):231--245.

\bibitem[Song et~al., 2021]{song2021visualizing}
Song, Y., Ren, S., Wolfson, J., Zhang, Y., Brown, R., and Fan, Y. (2021).
\newblock Visualizing, clustering, and characterizing activity-trip sequences via weighted sequence alignment and functional data analysis.
\newblock {\em Transportation Research Part C: Emerging Technologies}, 126:103007.

\bibitem[Tibshirani, 1996]{tibshirani1996regression}
Tibshirani, R. (1996).
\newblock Regression shrinkage and selection via the lasso.
\newblock {\em Journal of the Royal Statistical Society Series B: Statistical Methodology}, 58(1):267--288.

\bibitem[Van~de Geer et~al., 2011]{van2011adaptive}
Van~de Geer, S., B{\"u}hlmann, P., and Zhou, S. (2011).
\newblock The adaptive and the thresholded lasso for potentially misspecified models (and a lower bound for the lasso).

\bibitem[{Vincent} et~al., 2014]{msgl_csda}
{Vincent}, {M.}, {Hansen}, and R., N. (2014).
\newblock Sparse group lasso and high dimensional multinomial classification.
\newblock {\em Computational Statistics \& Data Analysis}, 71:771--786.

\bibitem[{Vincent} et~al., 2019]{msgl_manual}
{Vincent}, {M.}, {Hansen}, and R., N. (2019).
\newblock {\em msgl: Multinomial sparse group lasso}.
\newblock R package version 2.3.9.

\bibitem[Yuan and Lin, 2006]{grouplasso}
Yuan, M. and Lin, Y. (2006).
\newblock Model selection and estimation in regression with grouped variables.
\newblock {\em Journal of the Royal Statistical Society Series B: Statistical Methodology}, 68(1):49--67.

\bibitem[Zhao and Yu, 2006]{zhao2006model}
Zhao, P. and Yu, B. (2006).
\newblock On model selection consistency of lasso.
\newblock {\em Journal of Machine learning research}, 7(Nov):2541--2563.

\bibitem[Zhou, 2010]{zhou2010thresholded}
Zhou, S. (2010).
\newblock Thresholded lasso for high dimensional variable selection and statistical estimation.
\newblock {\em arXiv preprint arXiv:1002.1583}.

\bibitem[Zou, 2006]{zou2006adaptive}
Zou, H. (2006).
\newblock The adaptive lasso and its oracle properties.
\newblock {\em Journal of the American statistical association}, 101(476):1418--1429.

\end{thebibliography}

\newpage

\begin{center}
\section*{Supplemental Materials}
\subsubsection*{Multinomial thresholded LASSO for interpretable dimension reduction of human activity sequences}
\end{center}

\setcounter{figure}{0}
\renewcommand{\thefigure}{S\arabic{figure}}

\setcounter{table}{-1}
\renewcommand{\thetable}{S\arabic{table}}

\section*{I: Cluster assignment construction}

\vspace{7mm}

{\setstretch{1.4}

\framebox{\parbox{\dimexpr\linewidth-2\fboxsep-2\fboxrule}{\itshape
{\bf{Process for clustering sequences:}}
\begin{enumerate}
    \item Calculate pairwise distance matrix using optimal matching; perform hierarchical clustering with average linkage from the distance matrix.
    \item Cut the hierarchical clustering dendrogram with various pre-specified numbers of final clusters (e.g. integers between 2 and 100)
    \item Evaluate cluster assignments based on different numbers of clusters using the Dunn Index. Choose the number of clusters with optimal performance.
    \item Remove very small clusters as ``noise" or group them together. In this case, 23 clusters are identified. Clusters with size smaller than 13 (less than 0.7\% of the total size) are grouped together into Cluster 15.
\end{enumerate}
}}

}

\end{document}